\documentclass[conference]{IEEEtran}
\IEEEoverridecommandlockouts
\pdfoutput=1
\usepackage{cite}
\usepackage{amsmath,amssymb,amsfonts}
\usepackage{algorithmic}
\usepackage{graphicx}
\usepackage{textcomp}
\usepackage{xcolor}
\usepackage{authblk}
\usepackage[nolist]{acronym}
\usepackage[utf8]{inputenc}
\usepackage{url}
\usepackage{breakurl}
\usepackage[breaklinks]{hyperref}
\usepackage{tikz}
\usetikzlibrary{automata, arrows}

\def\BibTeX{{\rm B\kern-.05em{\sc i\kern-.025em b}\kern-.08em
    T\kern-.1667em\lower.7ex\hbox{E}\kern-.125emX}}
\begin{document}

\title{Confidential Computing for Privacy-Preserving Contact Tracing}

\author[1]{David Sturzenegger}
\author[ ]{Aetienne Sardon}
\author[1]{Stefan Deml}
\author[2]{Thomas Hardjono}
\affil[1]{decentriq}
\affil[2]{MIT Connection Science}
\affil[ ]{\textit {david.sturzenegger@decentriq.ch, aetienne.sardon@gmail.com, stefan.deml@decentriq.ch, hardjono@mit.edu}}
%\eid{123@gmail.com}

\maketitle

\begin{acronym}[aligator]
    \acro{BLE}{\textit{Bluetooth Low Energy}}
    \acro{TEE}{\textit{Trusted Execution Environment}}
    \acro{TTP}{\textit{Trusted Third Party}}
    \acro{SGX}{\textit{Software Guard Extensions}}
\end{acronym}

\begin{abstract}
Contact tracing is paramount to fighting the pandemic but it comes with legitimate privacy concerns. This paper proposes a system enabling both, contact tracing and data privacy.

We propose the use of the Intel \ac{SGX} trusted execution environment to build a privacy-preserving contact tracing backend. While the concept of a confidential computing backend proposed in this paper can be combined with any existing contact tracing smartphone application, we describe a full contact tracing system for demonstration purposes. 

A prototype of a privacy-preserving contact tracing system based on \ac{SGX} has been implemented by the authors in a hackathon.
\end{abstract}

\section{Introduction}
\label{sec:introduction}
The COVID-19 pandemic has caused a severe human and economic tragedy. As of April 2020, more than 1.2 million COVID-19 infections have been confirmed globally \cite{WHO}. Governments all over the world have taken action to prevent their health systems from being overwhelmed. National lockdowns as well as social and physical distancing measures have been imposed to slow the spread of disease, \cite{WHO2}. However, these measures have also brought large parts of the economy to a standstill, elevating the risk of a sustained economic downturn, \cite{Faulconbridge}.

\subsection{The importance of contact tracing}

Current research suggests that contact tracing could play a critical role in avoiding or leaving lockdown, \cite{Pentland}. Contact tracing can help maintain a relatively unrestricted society and economy, while minimizing the damage to the health of the population, \cite{Salathe}. Since most transmissions are estimated to occur from pre-symptomatic individuals, traditional manual contact tracing procedures are too slow to effectively contain the COVID-19 spread, \cite{Ferretti}. But smartphone apps that immediately alert recent close contacts and prompt them to self-isolate may significantly increase the efficacy of contact tracing.

\subsection{Privacy concerns}

It is estimated that 60\% of a country's population would need to participate in contact tracing for it to be effective, but privacy concerns may slow adoption \cite{Reuters}. The fundamental problem is the simple fact that to determine whether two people were in contact their location data needs to be compared. This directly conflicts with the desire of most people to keep their location data private, leading to a trade-off between health and privacy.

Contact tracing systems built by the Chinese and South Korean governments have favored health over privacy in the context of the current pandemic. These systems recently came under public scrutiny over issues of data protection and privacy. Critics argue that emergency measures  tend to be expanded beyond their original scope, \cite{TNW}. Hence, liberal countries are clearly in favor of opt-in based apps that use privacy-preserving technologies to minimize privacy and civil liberty intrusions, \cite{FT}. For example, a group of European experts recently launched the Pan-European Privacy Preserving Proximity Tracing Initiative to guide on best practices for developing contact tracing apps, but privacy concerns remain, \cite{PEPP}, \cite{Sieben}.

\subsection{From TPPs to TEEs}

Conventional systems rely on a \ac{TTP} to keep track of potential infection chains and orchestrate notifications (see section \ref{sec:current_status}). This has led some to conclude the need for elaborate governance structures. For example, in \cite{Dubach} the authors suggest amending the Epidemics Act to incorporate so called \textit{data trustees}, who shall be entrusted with guaranteeing proper data handling. Such considerations are based on the assumption that contact tracing requires the presence of a \ac{TTP}. However, with the advent of confidential computing this assumption seems outdated, as \acp{TEE} may make \acp{TTP} obsolete.

\subsection{Confidential computing}

Confidential computing refers to performing computations with additional data confidentiality and integrity guarantees. \acp{TEE} have recently emerged as one of the most flexible and mature technologies, which can enable confidential computing. Many of today's leading technology companies are actively developing and promoting confidential computing technologies. For example, companies like Microsoft, Google, Alibaba Cloud and others have joined forces to form the Confidential Computing Consortium under the Linux Foundation, \cite{CCC}. Currently, Intel's \ac{SGX} is the most advanced TEE implementation and the main technology the members of the Confidential Computing Consortium focus on, \cite{Intel}.

This paper proposes an Intel SGX-based contact tracing system which provably cannot reveal any user's location data while providing all benefits of a traditional contact tracing system. We focus on a confidential computing backend that can be used in combination with any of the currently existing contact tracing apps, requiring only minimal modifications.

\section{Current Contact Tracing Systems}
\label{sec:current_status}
Current contact tracing apps typically rely on pushing the infected user’s location data to the entire system. These location data include GPS and/or \textit{proximity data}, i.e. (typically randomized) identifiers of devices that were close to the current device. On every user’s phone, all infected users’ data are then compared to the locally stored location data in order to determine whether the mobile user has been within close proximity to infected individuals.

If the data is GPS data, this immediately reveals the infected user's past movements and offers very little privacy. If the data is proximity data, this may substantially leak privacy as well. Privacy loss may occur (a) to the mobile-phone user, and/or (b) to the diagnosed patient.\footnote{For example, an attacker may re-identify an individual by matching reported identifiers with his or her individual contact log. Second, an attacker may reconstruct a user's location history by matching any officially revealed identifiers with maliciously collected ones, e.g. through Bluetooth sniffing.} Attackers will prefer to attack large data sets located at Hospital servers.

These privacy problems can negatively influence a user's decision and willingness to disclose their infection to the system. They may therefore substantially degrade the system's overall effectiveness.

Two examples of existing contact tracing systems are discussed in the following.

\subsection{Example 1: HaMagen}
Israel's health ministry recently launched the contact tracing system HaMagen, \cite{MOH}, \cite{MOH2}, \cite{Moyal}. HaMagen claims that it only processes the users' location data on their devices. However, the system relies on pushing the location data of all infected users over government servers to all users in the system. Hence, the location data of infected people is not protected at all.

% HaMagen protects the data privacy of the healthy users and is a clear improvement over a central server collecting all data on an ongoing basis. However, it requires the infected users to share all their movement data with all users in the system. This sharing is likely to prevent some users from revealing their infection.

\subsection{Example 2: TraceTogether}
TraceTogether's approach is similar to the idea behind Apple’s ``find my device'' technology, \cite{SMOH}. Every active phone continuously monitors for \textit{Bluetooth Low Energy} (BLE) beacon messages, which are broadcasted from other devices together with some identifier. When it picks up one of these signals, the participating phone tags the data and stores it.

As a result, no location data is stored on device, but rather a list of ``identifiers'' of the users one has met. In order to make location tracking more difficult, regularly changing random identifiers, derived from a user's secret key, are used. However, in order to identify potential transmissions, an infected user has to reveal his or her entire proximity data to a central authority, increasing the risk of re-identification (see section \ref{sec:current_status}).

\section{Fundamental Problems}
\label{sec:fundamental_problems}
Current contact tracing apps need to address the following problems:
\begin{itemize}
\item \textbf{Revealing data of infected users.} Contact tracing apps like HaMagen perform on-device transmission detection. While this protects non-infected user data, it exposes infected individuals to re-identification risk by pushing their identifiers to all edge devices for local matching.
\item \textbf{Trustworthiness of central data processing.} Other systems, like TraceTogether or Pepp-PT \cite{Sieben}, require all edge devices to send their collected location or proximity data to a central server, where matching of infected and non-infected identifiers is performed. Typically, this makes it difficult for users to verify how their data is processed by the server. More specifically, it becomes impossible to guarantee that their contributed data will not be used beyond the pre-agreed purpose of contact tracing and will be deleted afterwards.
%\item \textbf{User fragmentation.} As mentioned in section \ref{sec:introduction}, the effectiveness of a contact tracing app greatly depends on its widespread adoption. But as innovators crowd the space, there is a growing risk of fragmenting the user base across different services, rendering them useless, \cite{FT}.
\end{itemize}

\section{Proposed Confidential Contact Tracing System}

Contact tracing systems consist of two components: the smartphone contact tracing app, installed on the user's device, and a contact tracing backend. While special-purpose \acp{TEE} exist on smartphones, they currently do not offer all the guarantees that are needed to conduct confidential computing. Especially the concept of remote attestation is lacking in most existing smartphone implementations, which makes them impractical for the use-case discussed in this work.

Hence we propose to build a confidential contact tracing backend to address the problems mentioned in section \ref{sec:fundamental_problems}. While this backend in general can be used with any contact tracing app, we propose a full contact tracing system (i.e. including a specific app) for demonstration purposes. 

The proposed backend shall leverage Intel \ac{SGX} to confidentially determine potential chains of transmission, without ever exposing any user data to anyone---not even to the platform operator. Much of the following description will not be particular to a confidential computing solution. The key benefits of using Intel \ac{SGX} are twofold: one can \textit{prove} that the system works as described, thereby preventing data misuse, and one can achieve a higher level of privacy protection than with conventional systems. 

\subsection{A Note on Software Guard Extensions}

Using \ac{SGX} technology, the GPS data from the infected patients are encrypted by the hospital in such a way that it can only be decrypted inside the \ac{SGX} \ac{TEE}. Similarly, GPS data from the user’s mobile phone is encrypted for the same target \ac{SGX} environment. Once both data sets are now within the trusted boundary of the \ac{SGX} \ac{TEE}, they can be decrypted safely and be compared. If a positive match is found, the \ac{SGX} \ac{TEE} will report the result over a secure channel to (a) the mobile user, and (b) optionally also to the hospital.

The benefit of the \ac{SGX} \ac{TEE} is that GPS data is never accessible in plaintext. Once the \ac{SGX} \ac{TEE} finishes the comparison of GPS data-sets, \ac{SGX} will delete (``flush'') the data from its memory. This ensures that the original GPS data-sets are present inside the \ac{SGX} \ac{TEE} only for a very short time. This has the advantage that attackers are unable to obtain access to large GPS sets. We therefore recommend that hospitals who are in possession of GPS data-sets of infected patients to encrypt their data while in storage.

\subsection{High-Level System Description}

\begin{figure}
\begin{center}
\begin{tikzpicture}[>=stealth',shorten >=1pt,auto,node distance=4cm]
  \node[state] (a) {A};
  \node[state] (c) [below right of=a, yshift=-0.8cm] {C};
  \node[state] (b) [above right of=c, yshift=0.8cm] {B};
  %\node[state, fill=gray] (h) [below of=c, yshift=2.4cm] {H};

  \path[->]          (a)  edge   [bend left=15]   node {$a_1$} (b);
  \path[->]          (b)  edge   [bend left=15]   node {$b_1$} (a);
  \path[->]          (a)  edge   [bend left=15]   node {$a_1$} (c);
  \path[->]          (c)  edge   [bend left=15]   node {$c_1$} (a);
  \path[->]          (b)  edge   [bend left=15]   node {$b_2$} (c);
  \path[->]          (c)  edge   [bend left=15]   node {$c_2$} (b);
  %\path[->]          (c)  edge                 node [swap] {$s_t$} (h);
\end{tikzpicture}
\caption{Example of random identifier communication flows.}
\label{fig:random_identifier_flow}
\end{center}
\end{figure}
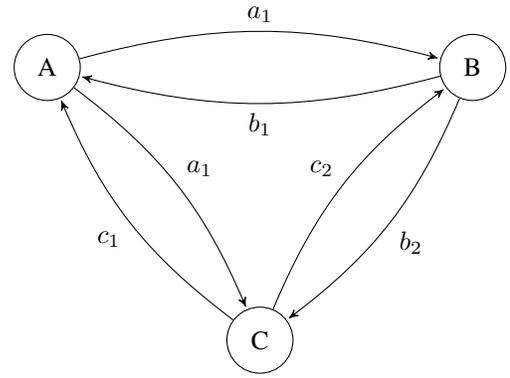

Assume each device generates and emits a random identifier in discrete time intervals $\Delta t$. For example, device $A$ emits $a_1$ during $[t_0 + \Delta t)$ and $a_2$ during $[t_0 + 2\Delta t)$ and so forth. Devices in proximity\footnote{``Proximity'' can potentially be determined through additional technology.} to one another pick up these random identifiers reciprocally and locally store the sent and received identifiers in a contact tuple log. For example, assume that, while in proximity to one another, device $A$ sends $a_1$ and device $B$ sends $b_1$. In this case, $A$ locally stores $(a_1, b_1)$ and $B$ stores $(b_1, a_1)$ (see figure \ref{fig:random_identifier_flow}).

Let's now assume user $C$ is tested at a health authority $H$. If the test is positive, the health authority submits this information to the confidential computing backend. 

The (authenticated) user $C$ polls the backend to see whether or not the test was positive. Note that $C$ cannot produce false infection notifications, since only the health authority can perform these calls to the backend. 

If $C$ decides to notify the user network of his infection, he or she sends his contact tuple log, e.g. $\{(c_{1},a_1),(c_{2},b_2),...\}$, to the backend which stores it in an encrypted database that is provably\footnote{\label{footnote1}Note that giving this proof is only possible using confidential computing.} only accessible to the backend\footnote{Alternatively, infected people could upload a secret which allows deriving their identifiers for any given point in time.}. 

All devices regularly poll the backend for matches in the encrypted storage by sending their contact tuples. For example, when $A$ polls the storage, he or she sends $\{...,(a_1, c_{1}),...\}$. As there was a match between, $A$'s and $C$'s tuples, $A$ is informed that he has been in contact with an infected individual. Note that provably neither $A$'s tuples nor the fact that $A$ was in contact with an infected individual get stored or submitted elsewhere by the backend (consider again footnote \ref{footnote1}).

\section{Discussion}

Much of the system proposed above is similar to existing contact tracing systems. The main difference consists of the fact that using Intel \ac{SGX}, it can be proved using \textit{remote attestation} as well as \textit{memory encryption} and \textit{memory isolation} (see also \cite{dq-blog} for a high-level introduction to these concepts) that the backend operates exactly as advertised. It is important to stress again that the concept of a confidential computing backend can be used in combination with \textit{any} contact tracing application, not just the one described here for demonstration purposes.

A prototype of a privacy-preserving contact tracing system based on \ac{SGX} has been implemented and open-sourced in the context \textit{CodeVsCovid19} hackathon, \cite{codevscovid19}.

The main benefits of the confidential computing-based backend are twofold: On the one hand it enables effective data minimization (i.e., data does not need to be exposed to perform contact tracing logic);\footnote{Note that it would also be possible to design a peer-to-peer based contact tracing system, in which infected users could establish secure communication channels directly with other users with whom they were in contact. However, in such a system it would be difficult to guarantee that the information of potential chains of transmission is propagated completely throughout the system, e.g., a node could refuse to forward a potentially infectious contact, undermining $n-$order notifications.} and, on the other hand, it provides transparent and verifiable data processing. This means that users can be guaranteed that their data is only used for the pre-agreed specific purpose of contact tracing. The specific data processing logic can be open-sourced, audited and verified through independent parties.

Note that the confidentiality and integrity guarantees of any system---including confidential computing systems---depend on a correct implementation. We did not describe such a full implementation. The purpose of this paper is to demonstrate the concept and feasibility of a privacy-preserving contact tracing system.   

We believe that a privacy-preserving backend can enable a more widespread and therefore effective contact tracing system.

\section{Conclusion}
We described the need for a privacy-preserving contact tracing solution: Without a strong focus on data privacy, contact tracing is unlikely to be widely adopted in liberal countries. We propose the use of Intel \ac{SGX} to build a confidential computing backend that provably cannot reveal any user data and outline a complete contact tracing system for demonstration purposes. Together with currently available contact tracing smartphone applications, such a privacy-preserving contact tracing system could help mitigate some of the adverse effects of the current pandemic. 

\bibliographystyle{IEEEtran} \bibliography{IEEEabrv,cocotrace}

\end{document}